# No clear evidence of ferroelectric order in tensile-strained anatase-TiO$_2$ thin films grown on (110) NdGaO$_3$ substrates

*Stella Skiadopoulou, Stanislav Kamba,* Jan Drahokoupil, Jan Kroupa, Nitin Deepak, Martyn E. Pemble and Roger W. Whatmore*

S. Skiadopoulou, Dr. S. Kamba, Dr. J. Drahokoupil, Dr. J. Kroupa
Institute of Physics, The Czech Academy of Sciences
Na Slovance 2, 18221 Prague 8, Czech Republic
E-mail: stella@fzu.cz; kamba@fzu.cz

N. Deepak, Prof. M.E. Pemble, Prof. R.W. Whatmore
Tyndall National Institute
University College Cork, "Lee Maltings"
Dyke Parade, Cork, Ireland

N. Deepak, Prof. M.E. Pemble, Prof. R.W. Whatmore
Department of Chemistry, University College Cork, Dyke Parade, Cork, Ireland

Prof. R.W. Whatmore
Department of Materials, Imperial College London, South Kensington Campus, London SW7 2AZ, UK

**Abstract**

Very recently there was a report of the discovery using piezoelectric force microscopy (PFM) of a switchable ferroelectric polarization in 1.6 % tensile strained TiO$_2$ thin film with anatase crystal structure (see N. Deepak et al. *Adv. Funct. Mater*. 2014, *24*, 2844). The polarization disappeared only above 450 K, which was assigned as a Curie temperature, $T_C$. Here we have performed X-ray diffraction, second-harmonic generation (SHG) and infrared investigations of the same films. Phonon frequencies exhibit less than a 10 % shift down with the tensile strain and no anomaly near expected $T_C$. SHG experiment did not reveal any signal characteristic for inversion symmetry breaking, as expected in ferroelectric phase, and the *c*-lattice parameter exhibits no anomaly on heating near expected $T_C$. Based on these results, we can summarize that we were not able to confirm the previously discovered ferroelectricity in tensile-strained anatase-TiO$_2$ thin films.

# 1. Introduction

The extensive diversity of photoelectric, photochemical, catalytic and dielectric properties of titanium dioxide ($TiO_2$) has led to a wide range of applications in traditional industry (e.g. white pigment, anti-corrosion coating, photocatalyst)[1,2] and the ground-breaking field of microelectronics (e.g. dielectric gates, optoelectronic and electrochromic devices). In addition, the reported, but still controversial, high-temperature ferromagnetism in undoped[3] and magnetically doped[4,5] $TiO_2$ is extremely attractive for the field of spintronics.

Among the main $TiO_2$ structures: the two tetragonal, rutile and anatase, the orthorhombic brookite and the monoclinic akaogiite,[6,7] rutile is the most studied. It is characterized by incipient ferroelectricity,[8,9,10] with a dramatic increase of the dielectric constant with decreasing temperature and absence of the ferroelectric phase transition even near 0 K. Interestingly enough for the ferroelectrics community, a series of theoretical studies have predicted the appearance of strain-induced spontaneous polarization in bulk rutile structures[11,12,13,14,15] and thin films.[16] Nevertheless, no experimental evidence of ferroelectricity has been reported as-yet for the rutile phase of $TiO_2$.

The only observation of any possible experimental evidence for the existence of ferroelectric behavior connected to any of the $TiO_2$ polymorphs was published very recently from Deepak at al.[17], which was for epitaxially strained anatase $TiO_2$ (a-$TiO_2$) thin films. A polar piezoelectric domain structure and electrically switchable polarization were observed by the use of piezoresponse force microscopy (PFM), for film thicknesses varying from 20 to 100 nm and reported as possible evidence for ferroelectric behavior. The polarization was stable against prolonged exposure to elevated temperature, implying a non-electrochemical origin for the effects observed, but disappeared on heating above 180 to 190°C, implying the possible existence of a Curie point, $T_C$. The films were grown on 0.5 mm – thick (110)

NdGaO$_3$ (NGO) substrate via liquid injection chemical vapor deposition (LI - CVD). The induced tensile strains were approximately 1.67 %, 1.44 % and compressive -1.57 % for the *a*-, *b*- and *c*-axis respectively for the case of 20 nm – thick films, and 0.7 %, 0.5 % and -0.4 % for the 100 nm – thick films. The ferroelectric polarization was observed out-of the film plane, i.e. along *c*-axis.[17]

In the current work, we report the lattice vibrations investigation via IR spectroscopy of the same a-TiO$_2$/NGO thin films. No substantial phonon-frequency shift expected near the temperature of the strain-induced displacive ferroelectric phase transition was observed. Additionally, X-ray diffraction studies of the out-of-plane lattice parameter as a function of temperature showed no anomaly close to the suspected Curie temperature of about 460 K. Room-temperature second harmonic generation (SHG) measurements did not reveal any appreciable signal above noise level, although this technique is sensitive to the breaking of space inversion symmetry in improper ferroelectrics with a polarization three orders of magnitude smaller than in canonical ferroelectrics BaTiO$_3$. On the basis of this evidence, we cannot support the proposition that strained anatase thin films exhibit ferroelectricity at room temperature, and conclude that a-TiO$_2$/NGO is paraelectric at ambient and higher temperatures.

## 2. Results and Discussion

### 2.1. Phonon studies

Far IR spectroscopy measurements were conducted, in order to investigate the phonon frequencies dependence on the film strain. Three thicknesses of a-TiO$_2$/NGO thin films were measured: 20, 80 and 100 nm. The measurements were carried out with a polarized beam and two different orientations with respect to the crystallographic axis of the NGO substrate: ***E*** || [110] and ***E*** || [001]. **Figure 1** shows the room temperature far IR reflectivity spectra of the

NGO substrate together with the far IR reflectance spectra of the a-TiO$_2$/NGO thin films, for the two different polarization orientations. The calculated complex permittivity from the experimental IR reflectivity spectra (as described below) is also presented for the a-TiO$_2$/NGO thin films. The corresponding complex permittivity spectra of a-TiO$_2$ single crystal, reported by Gonzalez *et al.*,[18] are included for comparison reasons.

A model corresponding to a two-layer optical system was used for the evaluation of the spectra.[19] The IR reflectivity spectra of the bare substrate were fit first. The reflectivity $R(\omega)$ is related to the complex dielectric function $\varepsilon^*(\omega)$ by

$$R(\omega) = \left| \frac{\sqrt{\varepsilon^*} - 1}{\sqrt{\varepsilon^*} + 1} \right|^2 \tag{1}$$

The complex permittivity $\varepsilon^*(\omega)$ of the NGO substrate was described by a generalized, factorized damped harmonic oscillator model[20]

$$\varepsilon^*(\omega) = \varepsilon_\infty \prod_j \frac{\omega_{LOj}^2 - \omega^2 + i\omega\gamma_{LOj}}{\omega_{TOj}^2 - \omega^2 + i\omega\gamma_{TOj}} \tag{2}$$

where $\omega_{TOj}$ and $\omega_{LOj}$ are the frequencies of the *j*-th transverse optical (TO) and longitudinal optical (LO) phonons, and $\gamma_{TOj}$ and $\gamma_{LOj}$ are the corresponding damping constants. $\varepsilon_\infty$ is the high-frequency (electronic) contribution to the permittivity, determined from the room-temperature frequency-independent reflectivity tail above the phonon frequencies. The obtained oscillator parameters of the substrate ($\omega_{TOj}$, $\omega_{LOj}$, $\gamma_{TOj}$, $\gamma_{LOj}$ and $\varepsilon_\infty$) were fixed for the fitting of the reflectance spectra of the a-TiO$_2$/NGO thin films. The model used to describe the complex permittivity of the thin films is a sum of *N* independent three-parameter damped harmonic oscillators, which can be expressed as[20]

$$\varepsilon^*(\omega) = \varepsilon_\infty + \sum_{j=1}^{N} \frac{\Delta\varepsilon_j \omega_{TOj}^2}{\omega_{TOj}^2 - \omega^2 + i\omega\gamma_{TOj}} \tag{3}$$

where $\Delta\varepsilon_j$ is the dielectric strength of the *j*-th mode.

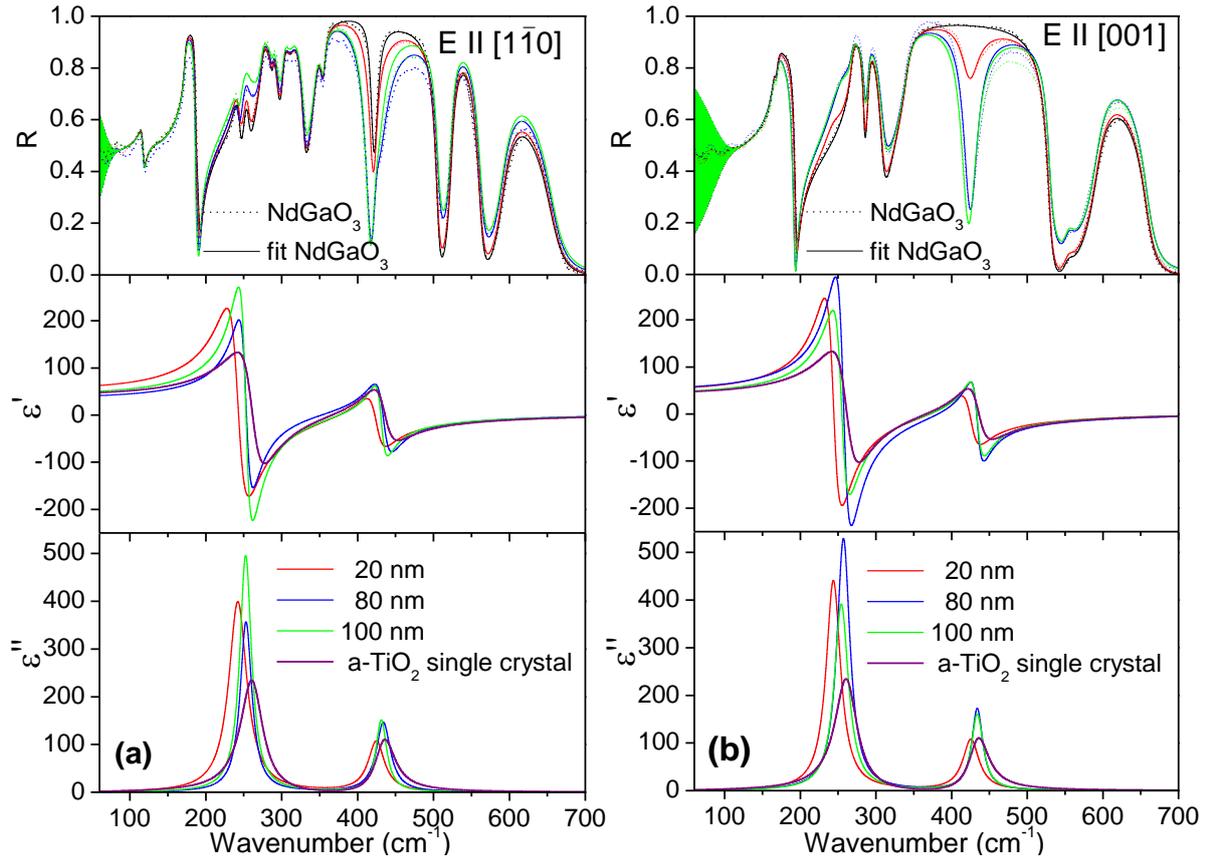

**Figure 1.** IR reflectance spectra and calculated complex permittivity of a-TiO$_2$/NGO thin films, measured with a polarized beam with: (a) ***E*** ∥ [1-10] and (b) ***E*** ∥ [001], with respect to the crystallographic axis of the substrate. The thin film IR reflectance spectra are compared with the IR reflectivity of the bare NGO substrate. Dotted lines correspond to the experimental data and the continuous to the fitted ones. For comparison, the complex permittivity spectra of a-TiO$_2$ single crystal are presented, according to the published ***E***⊥*c* data from Gonzalez *et al.*.[18]

Two phonon modes were detected for all the films and both polarization orientations, in close agreement with the single crystal phonon parameters (two $E_u$ symmetry modes are active in ***E***⊥*c* polarized spectra of a-TiO$_2$ crystals) published by Gonzalez *et al.*.[18] The shift of the TO phonon frequencies can be seen in the IR reflectance spectra of the films in comparison with the single crystal, despite the strong substrate contribution to the reflectivity and its elevated number of phonon modes. The calculated TO phonon frequencies from the fits of the experimental spectra of the three different a-TiO$_2$/NGO thin films are presented in **Table 1**, together with the corresponding film strains, as published by Deepak *et al.*.[17] One can also see lower phonon damping in strained thin films than in single crystal. The latter gives evidence of the very good quality of the epitaxial thin films.

IR reflectance spectra were measured up to 675 K. No temperature shifts of phonon frequencies where observed, only damping of the modes increases typically on heating. It is demonstrated in $\varepsilon''(\omega)$ spectra (Fig. 2) by the broadening of the peaks, as calculated from the fits.

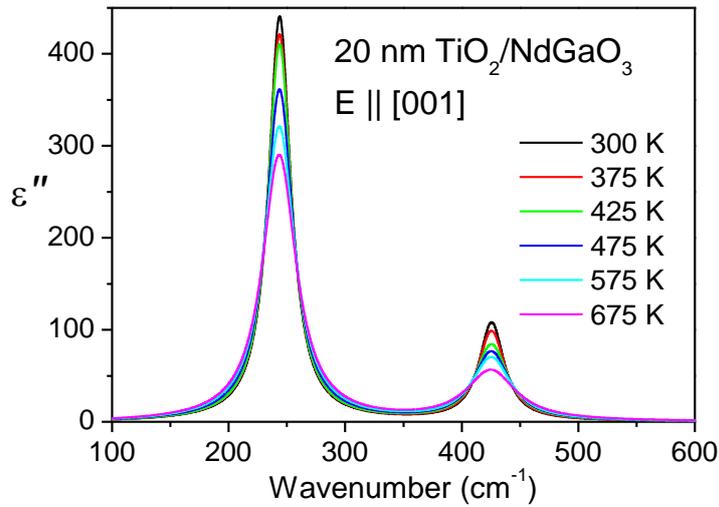

Figure 2. Dielectric loss spectra calculated from temperature-dependent $E \parallel [001]$ polarized reflectance spectra of 20 nm thin film of a-TiO$_2$.

The observation of a strain-induced ferroelectric phase transition has been published e.g. in SrTiO$_3$ and EuTiO$_3$ thin films.[21,22] In these particular cases, the tensile strain inducing ferroelectric polarization was lying in the film plane and due to the displacive type of the phase transitions, those were driven by a ferroelectric soft mode (i.e. polar optical phonon, which reduces its frequency near $T_C$) polarized in the film plane. [22,23]

**Table 1.** TO phonon frequencies of $E_u$ symmetry in the a-TiO$_2$/NGO films with different thicknesses and for the two different orientations of the polarized IR beam with respect to NGO crystal axes. The epitaxially-induced strain of the films for the three lattice axes[17] is listed. The single crystal TO phonon frequencies of $E_u$ symmetry[18] are presented for comparison.

| Thickness (nm) | Strain (%)[a] | | | $\omega_{TO1}$ (cm$^{-1}$) | | $\omega_{TO2}$ (cm$^{-1}$) | |
|---|---|---|---|---|---|---|---|
| | a-axis | b-axis | c-axis | **E** ‖ [110] | **E** ‖ [001] | **E** ‖ [110] | **E** ‖ [001] |
| Single crystal | - | - | - | 262.0[b] | 262.0[b] | 435.0[b] | 435.0[b] |
| 100 | 0.80 | 0.50 | -0.50 | 254.6 | 252.8 | 431.3 | 434.1 |
| 80 | 0.50 | 0.40 | -0.30 | 257.3 | 253.2 | 434.1 | 434.1 |
| 20 | 1.67 | 1.44 | -1.57 | 244.6 | 242.7 | 424.9 | 425.8 |

In the a-TiO$_2$ films, ferroelectric polarization was observed perpendicularly to the film plane, i.e. along $c$ crystal axis, i.e. surprisingly along the direction of compressive strain.[17] Therefore, the ferroelectric soft mode (if it exists) should be polarized along the same $c$ axis and will have $A_{2u}$ symmetry.[18] Unfortunately, the $A_{2u}$ symmetry phonons cannot be activated in IR reflectance spectra measured in near-normal reflection geometry used in our experiment. We can see only the **E**⊥$c$ response, i.e. $E_u$ symmetry modes. Nevertheless, if the $A_{2u}$ symmetry phonon would exhibit some instability near expected $T_C$ in strained a-TiO$_2$ films, the $E_u$ modes should also be strongly sensitive to the strain. In our case, we observed only a small phonon frequency shift of TO1 mode from 262 cm$^{-1}$ in crystal to 245 cm$^{-1}$ in the film with the highest strain of 1.67 % (20 nm – thick film). The shift of TO2 mode frequency was even smaller, namely from 435 cm$^{-1}$ to 425 cm$^{-1}$. For the two cases of the 80 and 100 nm – thick films, having strains of the order of ~ 0.5 %, the phonon frequency shifts are smaller and approximately the same within the experimental error – see Table 1. It is hard to imagine that the $A_u$ mode exhibits dramatic softening under strain, if the $E_u$ modes shift less than 7% with the strain. In single crystal, the $A_u$ frequency is 367 cm$^{-1}$ [18] at room temperature, which is too high for ferroelectric soft mode. Moreover, if this mode will soften under tensile strain, the $E_u$ modes should harden with the strain, whereas we see the opposite behavior. Thus, our IR spectra do not indicate any tendency to strain-induced lattice instability in a-TiO$_2$ films, which should be expected at displacive ferroelectric phase transition. If the a-TiO$_2$ film is ferroelectric, it can only belong to a family of materials with the phase transition of order-disorder type, where the ferroelectric phase transition is driven by a relaxation soft mode[24] or it can be a so-called improper ferroelectric.[25] In the latter case, the order parameter is not polarization, so no polar soft mode drives the phase transition and a negligible dielectric anomaly can be expected at $T_C$.

---
[a] Taken from ref. [17]; [b] Taken from ref. [18].

In rutile, the $A_{2u}$ mode exhibits much lower frequency (172 cm$^{-1}$ at 300 K) than in anatase, therefore its tendency to softening on cooling is remarkable.[26] For that reason rutile exhibits incipient ferroelectric behavior at low temperatures[8] and theoretical calculations[15,16] predicting its softening under strain are plausible. Unfortunately, IR and dielectric studies of strained rutile film are still missing.

**2.2. X-ray Diffraction**

At the ferroelectric phase transition the crystal structure changes and some anomalies (kinks or splitting) in temperature dependence of the lattice constants are expected. In the thin films, the measurements of all lattice parameters are complicated, but the out-of-plane lattice parameter can be easily and precisely measured. Using this method, dramatic shifts of the phase transition temperatures were revealed in strained PbTiO3 and BaTiO3 ultra-thin films.[27] Figure 3 shows the temperature dependence of c parameter (note that expected ferroelectric polarization should be oriented along c direction)[17] in 20 nm-thick a-TiO2 film. It exhibits a classical thermal expansion and within our accuracy of measurements no anomaly near expected TC ≈ 450 K. Therefore, our X-ray diffraction does not reveal any signature of ferroelectric or structural phase transition between 300 and 620 K.

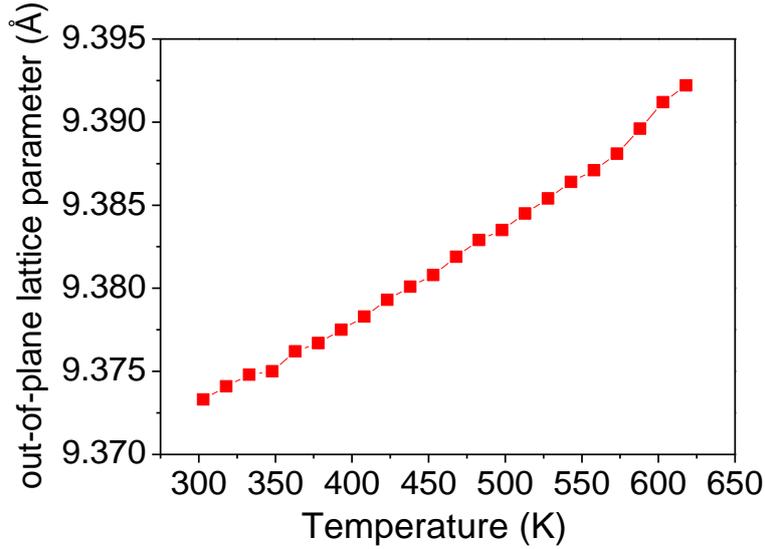

**Figure 3.** Out-of-plane lattice parameter of a-TiO$_2$/NGO as a function of temperature, as measured by X-ray diffraction, from room temperature up to 618 K. Error bars have the same size as the symbols. No anomaly near expected $T_C$ = 450 K is seen.

**2.3. SHG studies**

A true ferroelectric phase has broken space symmetry and therefore should give rise to a detectable SHG signal. For example, the SHG technique was used for detection of the strain-induced shift of the ferroelectric phase transition in BaTiO$_3$/GdScO$_3$.[27] This technique is very sensitive to acentric crystal structure even in polar nanoclusters in strained Sr$_{n+1}$Ti$_n$O$_3$ (n=1-6,∞) thin films[28] or relaxor ferroelectrics,[29] and so if a-TiO$_2$/NGO were ferroelectric at room temperature, an SHG signal should be detected. In the 20 nm and 100 nm thick films we did not detect any SHG signal above the noise level. For comparison, we also measured the SHG signal from ferroelectric BaTiO$_3$ thin film (thickness 25 nm) grown on (110) GdScO$_3$ using molecular beam epitaxy in the group of Prof. Schlom from Cornell University and it was easy to observe an SHG signal with signal-to-noise ratio better than 20 dB. In this case, the detection limit of our apparatus was about 2 orders of magnitude below the BaTiO$_3$ film signal. We note that the SHG technique was able to detect broken space symmetry even in 6 nm thin TbMnO$_3$ film,[30] where the spin-induced (i.e. improper) ferroelectric polarization is 500-times smaller than in BaTiO$_3$. The absence of SHG signal in our experiments does not support acentric ferroelectric structure of our a-TiO$_2$ thin films.

## 3. Discussion and conclusion

Our infrared reflectance studies performed on the same films as in Ref. [17], revealed no softening of $E_u$ phonons with strain and heating, so the hypothetical phase transition cannot be a displacive ferroelectric one, even if we cannot directly measure the $A_{2u}$ symmetry soft phonon. If the material is ferroelectric, its phase transition can be only of order-disorder type, where the structural change is driven by a soft dielectric relaxation with relaxation frequency below optical phonon frequencies. X-ray diffraction measurements of the *c*-lattice parameter did not reveal any anomaly up to 620 K, typical for a structural phase transition. Moreover, room-temperature SHG measurements did not reveal any signal typical for broken inversion symmetry, although the SHG method is in principle sensitive on a small inversion-symmetry breaking even in spin-induced (i.e. improper) ferroelectric TbMnO$_3$ thin films[30] (thickness 6 nm) with polarization 500-times smaller than in BaTiO$_3$.

One can see above that our results do not support, within accuracy of our measurements, the existence of ferroelectric phase transition in tensile-strained a-TiO$_2$ thin films grown on NdGaO$_3$ substrates. One can speculate that the switchable ferroelectric-like response can be caused by electric-field induced migration of various defects (oxygen or Ti vacancies, Ti in interstitial positions etc. – see e.g. review of Szot et al.[31]). It was demonstrated e.g. in non-ferroelectric SrTiO$_3$ that bias-induced defect migration can give rise to an apparent polarization or piezoelectric hysteresis loops, which are difficult to distinguishable from true ferroelectric hysteresis loops.[31,32] Nevertheless, in this case the polarization should decay with time, but as reported in Ref. [17] the poled domains in a-TiO$_2$ were stable for more than 48 h at 300 K. Surprisingly, at 353 K the piezoresonse did not decay with time but even increased 3.5-times after 3 h and remained stable for more 5 h. This increase with time could be explained by the loss of surface moisture,[17] but it probably requires further studies in order to be sure. In any case, the lack of decay of polarization with time observed at elevated temperatures does not support the idea of defect migration in electric field. Only between 450 and 470 K the polarization disappeared, which was explained by a ferroelectric phase transition.[17]

One can claim that ferroelectric polarization is very small in a-$TiO_2$ films and therefore the expected SHG signal is too weak to be detected in the thin films. Here we can argue that the polarization in a-$TiO_2$ should be robust according to Ref. [17]. Based on our search in the literature, there is no report of piezoelectric hysteresis loops observed by the use of PFM in weak ferroelectrics, such as spin-order induced ferroelectrics (e.g. $TbMnO_3$). Therefore, the piezoresponse observed in Ref. [17] would be equivalent to a ferroelectric polarization of at least two orders of magnitude higher than in $TbMnO_3$. In such a case the SHG signal should have easily been detected.

We can speculate that the piezoresponse is caused by an electric-field-induced electrochemical process, which can be irreversible at room temperature and disappears only above 470 K. Such processes have been frequently seen in many systems – for a review see e.g. Jesse et al. [33] However, the existence of such a process in strained a-$TiO_2$ films cannot be confirmed.

Finally, we can summarize that our SHG, XRD and lattice dynamics studies did not reveal any hint of ferroelectric phase transition in strained a-$TiO_2$ film. There are two possible explanations: the ferroelectric distortion is too small (less than 0.05 $\mu C/cm^2$) and cannot be detected by the SHG and XRD techniques used, or the piezoelectric response reported in Ref. [17] is defect and/or electric field induced, in spite of the fact that it was stable with time at elevated temperatures.

## 4. Experimental Section

*Samples:* For our investigations we used the same biaxially-strained epitaxial thin films of a-$TiO_2$ grown on (110) NGO substrates like in Ref. [17]. Details of the thin film deposition and results of TEM and PFM characterizations can be found in Ref. [17].

*IR reflectivity:* IR reflectance measurements were performed using a Bruker 113v Fourier transform IR spectrometer. Pyroelectric detectors from deuterated triglicine sulfate were used for detection of IR radiation. The reflectance measurements were performed in near normal incidence polarized beam with polarizations **E** ∥ [001] and **E** ∥ [1$\bar{1}$0] with respect to the (110) $NdGaO_3$ substrate. A commercial high-temperature cell (SPECAC P/N 5850) was used for the high-temperature IR experiments.

*X-ray diffraction:* The XRD was measured on PANalytical diffractometer X`Pert PRO in parallel beam geometry. The Co (λ = 0.178901 nm) radiation was used. The diffractometer was equipped with temperature chamber HTK 2000. The sample was firstly oriented on tungsten temperature element using ω-scans and manual rotation around out-of-plane axis. The parallel beam geometry is not sensitive to sample displacement. However, for single crystal or epitaxial grown thin layers, small inclination of the sample caused by temperature changes can break the diffraction condition and result in disappearing of the observed diffraction. Moreover, the omega scan of the measured 004 diffraction shows several components. For these reasons, the diffraction maxima were measured using two-axis scans (ω, 2θ-ω) and the 2θ-position of 004 diffraction was evaluated as average value of 2θ-positions of 2θ-scans for 15 particular ω-positions.

*Second harmonic generation:* For the SHG study we used our standard setup with a Q-switched Nd-YAG laser (λ = 1064 nm, 20 Hz, ~ 0.5 mJ/pulse, 6 ns) as the light source. The filtered SH signal at 532 nm was detected by a photomultiplier and a boxcar averager.

**Acknowledgements**

This work was supported by the Czech Science Foundation (Project No. 15-08389S) and the European Union's Seventh Framework Programme for research, technological development and demonstration under grant agreement No. 607521. We thank Prof. D.G. Schlom for providing 25 nm thick film of $BaTiO_3$ grown on $GdScO_3$ substrate for reference SHG measurements.